\documentclass[preprint]{aastex}
\usepackage{natbib}

\shorttitle{A Search for Nitrogen Enriched Quasars in the SDSS EDR}
\shortauthors{Bentz \& Osmer}

\begin{document}
\title{A Search for Nitrogen Enriched Quasars in the Sloan Digital 
       Sky Survey Early Data Release}
\author{Misty C. Bentz and Patrick S. Osmer}
\affil{Department of Astronomy, The Ohio State University}
\affil{140 W. 18th Ave, Columbus, OH 43210-1173}
\email{bentz,posmer@astronomy.ohio-state.edu}

\begin{abstract}
A search for nitrogen-rich quasars in the Sloan Digital Sky Survey Early
Data Release (SDSS EDR) catalog has yielded 16 candidates, including
five with very prominent emission, but no cases with nitrogen emission
as strong as in Q0353-383.  The quasar Q0353-383 has long been known to
have extremely strong nitrogen intercombination lines at $\lambda$1486
and $\lambda$1750 \AA, implying an anomalously high nitrogen abundance
of $\sim$ 15 times solar.  It is still the only one of its kind known.
A preliminary search through the EDR using the observed property of the
weak \ion{C}{4} emission seen in Q0353-383 resulted in a sample of 23
objects with unusual emission or absorption-line properties, including
one very luminous $z \approx 2.5$ star-forming galaxy.  We present
descriptions, preliminary emission-line measurements, and spectra for
all the objects discussed here.
\end{abstract}


\section{Introduction}
Q0353-383 was first observed in 1976 as part of the Curtis Schmidt
survey \citep{1980ApJS...42..333O}.  Subsequent observations of this z =
1.96 quasar revealed that Q0353-383 was an unusual object, in that its
spectrum revealed prominent \ion{N}{3}], \ion{N}{4}] and \ion{N}{5}
emission lines and abnormally weak \ion{C}{3}] and \ion{C}{4} lines
compared to other quasars. Figure 1a shows the spectrum of Q0353-383
(Baldwin, private communication) compared to the SDSS composite spectrum
in Figure 1b \citep{2001AJ....122..549V}.  \citet{1980ApJ...237..666O}
concluded that Q0353-383 has an anomalously high N abundance due to
recent CNO processing in stars.  Building on Osmer's original work with
Q0353-383, \citet{2003ApJ...583..649B} obtained higher quality spectra
and added HST observations in the ultraviolet.  Using the latest models
of the broad line region, Baldwin et al. were able to both confirm and
refine the original conclusions made by Osmer.  Additionally, Baldwin et
al. conclude that Q0353-383 has a metallicity of at least 5-10 times
solar.  As shown by simulations \citep{1999AR}, this level of over
abundance is expected to occur near the end of an era of rapid metal
enrichment which can result in metallicities of as much as 10-20 times
solar.  The scarcity of objects like Q0353-383 may be an indication of
the amount of time a quasar spends in this state of extreme metal
enrichment before the gas supply is exhausted and the quasar becomes
inactive.

To date, Q0353-383 is the only object of its kind known, which brings up
two questions: (1) how many other such objects exist, and (2) what
percentage of the QSO population is similar to Q0353-383?  Until
recently, anwering these questions was difficult due to the relatively
small number of known quasars and the lack of available spectra with
good S/N.  However, the Sloan Digital Sky Survey (SDSS) is providing new
opportunities by working to compile, in one database, approximately
100,000 quality quasar spectra as it scans 10,000 deg$^2$ of the north
Galactic cap \citep{2000AJ....120.1579Y}.  Even though the survey is not
complete at this time, a search for similar nitrogen-rich QSOs can be
undertaken.

\section{Spectral Analysis}
The SDSS EDR \citep{2002AJ....123..485S} Quasar Catalog
\citep{2002AJ....123..567S} is composed of 3814 objects that cover an
area on the sky of 494 deg$^2$.  Three thousand of these objects were
discovered by the SDSS in the commissioning data using a multicolor
selection technique similar to the one described by
\citet{2002AJ....123.2945R}.  Each object has at least one emission line
with a full width half-maximum of more than 1000 km s$^{-1}$, a
luminosity brighter than $M_{i^*} = -23$ mag, and a reliable redshift
\citep{2002AJ....123..567S}.  The sample is not homogeneous, as there
were several modifications to the quasar selection algorithm during the
time the data was collected, and so the EDR quasar catalog is not
intended for statistical purposes.  However, there are still enough data
contained in the catalog to make the information mined there important.

The SDSS EDR database was searched for all quasars that lay within the
redshift range \(1.8 < z < 4.1\).  This range ensures that both
\ion{N}{4}] $\lambda$1486 and \ion{N}{3}] $\lambda$1750 will be in the
3800--9200 \AA\ region observed by the SDSS spectrograph.  A total of
1082 QSOs were found in this range, with the redshift distribution as
shown in Figure 2.  We used the redshift values as specified in the
Quasar Catalog \citep{2002AJ....123..567S} to correct for cosmological
expansion and place the spectra in a common rest frame.  Within each
spectrum, two 30 \AA\ sections, centered on \ion{N}{4}] $\lambda$1486
and \ion{N}{3}] $\lambda$1750, were cross-correlated with the rest-frame
spectrum of Q0353-383 using the IRAF\footnote{IRAF is distributed by the
National Optical Astronomy Observatories, which are operated by the
Association of Universities for Research in Astronomy, Inc., under
cooperative agreement with the NSF.} package {\it fxcor}. The
cross-correlation routine was good at picking out objects with
relatively strong nitrogen emission, but it was also adept at finding
noise.  Therefore, objects with high correlation coefficients ($>0.3$)
for either species of nitrogen were visually inspected in order to
verify the presence or absence of nitrogen emission.  All objects with
typical broad absorption line profiles were immediately discounted.

We detected sixteen candidate nitrogen-rich objects within our sample of
1082 quasars.  While the sixteen objects selected had a correlation
coefficient of at least 0.5 for either \ion{N}{3}] or \ion{N}{4}] (see
Table 2), five of the candidates have noticeably stronger nitrogen
emission than the others (SDSS J1308$-$0050, SDSS J1327+0035, SDSS
J1744+5351, SDSS J2326$-$0020, and SDSS J2336$-$0017; see Table 2), but
none have nitrogen emission from both species as strong as that seen in
Q0353-383.  The SDSS spectrum of each object, smoothed over five pixels,
is shown in Figure 3.  Five of the sixteen candidate quasars were known
previous to the SDSS; however, only one (SDSS J0034-0111, aka UM 259)
was noted as possibly having strong nitrogen emission
{\citep[]{1977ApJS...35..203M}.

\subsection{Description of Candidates}

Due to the faintness of the objects, the spectra are somewhat noisy, and
so the equivalent widths of emission lines in this paper should be taken
as guideline measurements, with a characteristic error of 0.5-1 \AA.
The measurements were made using a simple summing function and a
two-point interpolation of the continuum within IRAF (see Table 2).
None of the objects presented here have nitrogen lines as strong as in
Q0353-383.  Many of the candidates do not even have an obvious
\ion{N}{4}] emission line, but this could, in part, be due to the noise
at the blue edge of the spectra for the $z \approx 2$ quasars.

In the cases of the two quasars where Ly$\alpha$ emission is visible in
the EDR spectra (SDSS J0953+0037 and SDSS J1707+6443, Figures 3g and 3m,
respectively), the line profile is rather narrow with a well-separated
\ion{N}{5} emission line.  This is also true of Q0353-383.  A literature
search on the previously known quasars revealed that SDSS J0259-0020
(\citealt{1991AJ....102..461C}, Fig. 1; \citealt{1994ApJ...436..678O},
Fig.1) and SDSS J1238-0059 (\citealt{1991AJ....101.1121H}, Fig. 1) also
follow this trend.  However, SDSS J1442-0037
(\citealt{1991AJ....101.1121H}, Fig. 2m) has a somewhat broader
Ly$\alpha$ profile with nearly indistinguishable \ion{N}{5} emission,
and SDSS J0034-0111 (\citealt{1994ApJ...436..678O}, Fig. 2k) has a very
broad, blended Ly$\alpha$ and \ion{N}{5} line profile.

There appear to be narrow components as well as broad components to the
emission line profiles in several of the spectra.  In fact, the FWHM of
the \ion{C}{4} line ranges from $\sim 800$ km s$^{-1}$ (SDSS J0953+0037,
Fig. 3g) to $\sim 7000$ km s$^{-1}$ (SDSS J0250-0047 and SDSS
J2336-0017, Figs. 3d and 3q) within the sample of 16 quasars.  SDSS
J0953+0037 is extremely strange in that it appears to be a narrow-line
QSO with many intrinsic Ly$\alpha$ and \ion{C}{4} absorbers.  It also
happens to be the one case where an [\ion{O}{1}] night sky line fell
exactly at $\lambda$1750, and so it is impossible to determine if there
is \ion{N}{3}] emission using the SDSS spectrum.

Of additional interest, seven of the candidates have strong,
well-separated \ion{He}{2} $\lambda$1640 and \ion{O}{3}] $\lambda$1664
emission lines (see Table 3), which is another of the odd
characteristics of Q0353-383.  There is evidence for \ion{He}{2} and
\ion{O}{3}] emission in most of the other spectra, but the S/N are too
low for even preliminary measurements of these weak lines.

The overall continuum shapes of the candidates are also rather varied.
Three candidates have a relatively flat continuum, six have a slight
slope to the blue in the continuum that is shortward of the \ion{C}{3}]
emission line, and the remaining seven have a distinct slope to the blue
throughout the visible portion of their spectra.  About one-third of the
candidate quasars also seem to have absorption signatures, although only
one seems to have a broad absorption profile (SDSS J1327+0035).

While it appears that most of the candidates are clustered right around
$z \approx 2$, this is to be expected as the distribution of quasar
redshifts in the SDSS EDR peaks near this redshift.

\section{Other Findings}
As noted by \citet{1980ApJS...42..333O}, the \ion{C}{4}/(Ly$\alpha$ +
\ion{N}{5}) ratio of Q0353-383 is extremely small, a value of 0.07,
which is 4 to 5 times smaller than the average value of the other
quasars in the Curtis Schmidt survey.  Therefore, as a preliminary
search, those quasars in the SDSS EDR database with a (non-negative)
ratio of \ion{C}{4}/(Ly$\alpha$ + \ion{N}{5}) $< 0.1$ were examined.
The \ion{C}{4}/(Ly$\alpha$ + \ion{N}{5}) ratios were calculated using
the crude equivalent widths supplied by the EDR pipeline, which included
the effects of absorption and/or emission. Table 4 lists the properties
of the 23 QSOs matching these criteria, none of which were previously
known.  Interestingly, the two nitrogen-rich candidates SDSS J0953+0037
and SDSS J1707+6443 have redshifts greater than 2.3 (where Ly $\alpha$
shifts into the SDSS spectrograph window of 3800--9200 \AA ), and yet
they were not among the 23 quasars that were selected (see Figure 4 for
spectra).  This particular search technique was not very useful for our
purposes, but it did a nice job of collecting extremely peculiar
objects, with broad absorption lines, intrinsic \ion{C}{4} and
\ion{N}{5} absorption, strong narrow Ly $\alpha$ emission with
well-separated \ion{N}{5} emission, and even one high-redshift
star-forming galaxy.  Two objects (SDSS J1303+0020 and SDSS J1426-0021,
Figures 4k and 4p, respectively) show evidence for strong outflows with
P-Cygni (narrower than normal BALs) line profiles of \ion{N}{5},
\ion{C}{4}, and the Si blend at $\lambda$1400.  

\subsection{A Star-Forming Galaxy Candidate at z $\approx$ 2.5}
One of the objects uncovered in our search, SDSS J1432-0001, was
included in the EDR quasar catalog on account of its strong Ly$\alpha$
emission and UV brightness.  This object, however, seems to be a
starburst galaxy (see Figure 5 for comparison with the composite Lyman
break galaxy (LBG) spectrum from \citealt{2003ApJ...588...65S}).  There
is strong evidence for the following interstellar absorption features in
the spectrum of SDSS J1432-0001: \ion{Si}{2} $\lambda$1260, \ion{O}{1}
$\lambda$1302, \ion{Si}{2} $\lambda$1304, \ion{C}{2} $\lambda$1334,
\ion{Si}{4} $\lambda \lambda$1393, 1402, \ion{Si}{2} $\lambda$1526,
\ion{C}{4} $\lambda \lambda$1548, 1550, \ion{Fe}{2} $\lambda$1608,
\ion{Al}{2} $\lambda$1670, and \ion{Al}{3} $\lambda\lambda$ 1854, 1862
\citep{2003ApJ...588...65S,1991ApJS...77..119M}.  The continuum of SDSS
J1432-0001 is redder than that of the LBG composite, and the lines are
somewhat broader as well.  Also, there seems to be very little
intergalactic absorption immediately shortward of Ly$\alpha$.  However,
while slightly different, SDSS J1432-0001 has many of the same emission-
and absorption-line features of the LBG composite, and it has similar
colors to the LBGs of the \citet{2003ApJ...592..728S} LBG survey, with
$u^* - g^* = 2.03$ and $g^* - r^* = 0.6$.

SDSS J1432-0001 has an observed r* magnitude of 20.52, which is two
magnitudes brighter than the brightest LBG in the
\citet{2003ApJ...592..728S} survey (Q0201-oMD12 at $z = 2.567$ has an
${\cal R}_{AB}$ magnitude of 22.67), and four magnitudes brighter than
typical $z \approx 3$ LBGs, which have magnitudes of {$\cal R$}$_{AB} =
24-25.5$ mag \citep{2003ApJ...588...65S}.  SDSS imaging of SDSS J1432-0001
shows no evidence for a foreground lensing galaxy, although this cannot
be ruled out as the SDSS data are insufficient to determine the presence
of sub-arcsecond image splitting or the presence of a foreground lensing
system at $z \approx 1$.

If we assume, however, that the continuum emission at $\lambda$1500 \AA\
arises only from young O and B stars, we can estimate the star formation
rate (SFR) of the galaxy from the flux at $\lambda$1500\AA\ $\times
(1+z)$ (F$_{\lambda}$(1500), which was measured directly from the
calibrated spectrum and has a value of F$_{\lambda}(1500) \approx 2.2
\times 10^{-17}$ erg s$^{-1}$ cm$^{-2}$ \AA$^{-1}$.  The flux was
converted to luminosity assuming a cosmological model with $\Omega_M =
0.3$, $\Omega_{\Lambda} = 0.7$, and $h = 0.7$.  Assuming a Salpeter IMF
with mass limits of 0.1 to 100 M$_{\odot}$ and a 10$^8$ year old
continuous star formation model, \citet{Kennicutt1998} derives the relation
\begin{equation}
{\rm SFR\ M_{\odot}\ yr^{-1}} = 1.4 \times 10^{-28} L_{\nu}(\lambda1500)
\end{equation}
between the star formation rate and the luminosity density L$_{\nu}$ (in
ergs s$^{-1}$ Hz$^{-1}$) at $\lambda$ = 1500 \AA.  The implied SFR is
therefore $\sim 390$ M$_{\odot}$ yr$^{-1}$, without correction for
attenuation by dust.  ``Normal'' spiral galaxies have typical SFRs of
$0-5$ M$_{\odot}$ yr$^{-1}$ \citep{1997A&A...324..490B}, and the LBGs of
the \citet{2003ApJ...592..728S} survey have SFRs of $\sim 50-100
h^{-2}_{70}$ M$_{\odot}$ yr$^{-1}$ after an average factor of $\sim$ 7
attenuation by dust \citep{2003ApJ...588...65S}.  As previously
mentioned, the spectrum of SDSS J1432-0001 is somewhat redder than that
of the LBG composite, implying a large correction for dust, and so the
SFR could be as much as a few thousand M$_{\odot}$ yr$^{-1}$.  With a
SFR this large, it would be expected that there would be differences
between the spectrum of such an object and its much less active
counterparts.

In viewing the spectrum, it is obvious that this object could not be a
normal AGN.  However, there is the possibility that it is some
heretofore unknown form of broad absorption line (BAL) quasar.  The line
widths are broader in SDSS J1432-0001 than in the composite spectrum,
and the luminosity is much higher.  However, the widths of these lines
are not outside the limits allowed by star formation models such as
STARBURST99 \citep{Leitherer1999}, and such a high degree of similarity
to the composite spectrum would be remarkable for a BALQSO.  As the area
covered by Lyman break surveys is very small compared to the area
covered by Sloan - only 0.38 deg$^2$ in the $z \sim 3$ survey by
\citep{2003ApJ...592..728S} - it is highly likely that SDSS J1432-0001
is simply a very rare, extremely bright LBG with a high SFR.

\section{Conclusions}
We have searched 1082 quasars in the SDSS EDR catalog for nitrogen-rich
objects similar to Q0353-383.  Sixteen candidates were found, five of
which have stronger nitrogen emission than the others (SDSS
J1308$-$0050, SDSS J1327+0035, SDSS J1744+5351, SDSS J2326$-$0020, and
SDSS J2336$-$0017), but none with the strength of nitrogen emission from
both \ion{N}{4}] $\lambda 1486$ and \ion{N}{3}] $\lambda 1750$ that is
seen in Q0353-383. Additionally, we found 23 objects with unusual
emission or absorption-line properties in our search for objects with
weak \ion{C}{4} emission relative to Ly$\alpha$ + \ion{N}{5}, one of
which appears to be a LBG with an extremely high SFR of $\sim 390$
M$_{\odot}$ yr$^{-1}$, without corrections for dust.
 
Further observations will be necessary in order to determine
conclusively whether any of these QSOs do indeed have extreme nitrogen
abundances.  The best candidates for follow-up observations are
J1308-0050, J1327+0035, J1744+5351, J2326-0020, and J2336-0017, as they
have the most prominent nitrogen emission lines of the sample and show
evidence of both \ion{N}{3}] and \ion{N}{4}] emission.  Higher S/N data
would allow for more precise measuring of equivalent widths.
Additionally, the wavelength range of the spectra may be pushed further
into the blue, allowing for measurements of Ly$\alpha$ and \ion{N}{5} on
several of quasars with $z \approx 2$.  With better measurements of the
line intensities, especially in the seven objects with prominent
\ion{O}{3}] and \ion{He}{2} emission, estimates of the metallicities
could be made using several different line ratios, such as
\ion{N}{3}]$/$\ion{O}{3}], \ion{N}{3}]$/$\ion{C}{3}], and
\ion{N}{5}$/$\ion{He}{2}.  These metallicity estimates would allow us to
test the hypothesis that nitrogen-rich quasars are exhausting their fuel
supply and approaching the metallicities expected by numerical
simulations for black holes as they end their quasar phase.

Searches through subsequent SDSS databases will also be necessary to
statistically determine the fraction of quasars that fit into the
``nitrogen-loud'' category and to ascertain the portion of a quasar's
lifetime that is spent in this last, high-metallicity stage before the
gas supply is exhausted.  It is worth mentioning that any estimate of
the frequency of occurence of these high-metallicity objects may only be
a lower limit, however, as both \ion{N}{4}] $\lambda$1486 and
\ion{N}{3}] $\lambda$1750 are collisionally de-excited at densities
greater than $10^{11}\ \rm{cm^{-3}}$ \citep{1999quco.conf..409H}.  There
may also exist high-metallicity objects with only \ion{N}{5} visible in
their spectra, which would be overlooked in our search.

Of the SDSS EDR objects we searched, only three to five of these objects
could seriously be considered candidate ``nitrogen-loud'' quasars, or
$\sim 0.4\%$ of the sample of 1082 quasars.  The other 12 of the sample
of 1082 quasars, or $\sim 1.1\%$, would more appropriately be labeled
``nitrogen-salient'' objects in that they have noticeable nitrogen
emission, but not to the extent of Q0353-383.  ``Nitrogen-salient''
objects may turn out to be valuable as well in that they may help to map
out the rapid chemical evolution stages of quasars as the fuel supply is
processed through stars.  While the above statistics are very rough due
to the inhomogeneity of the EDR sample, they do give a guideline
measurement of the frequency of such quasars based on a factor of ten
larger sample size than the previous estimate by
\citet{1980ApJ...237..666O}. We plan to search the SDSS First Data
Release database \citep{dr1preprint}, using the strategies developed in
this paper, to improve these statistics.

\acknowledgements
We would like to thank Brad Peterson and Rick Pogge for their helpful
comments and suggestions, and an anonymous referee for very thorough
feedback.  Misty Bentz is supported by a graduate fellowship of The Ohio
State University.

Funding for the creation and distribution of the SDSS Archive has been
provided by the Alfred P. Sloan Foundation, the Participating
Institutions, the National Aeronautics and Space Administration, the
National Science Foundation, the U.S. Department of Energy, the Japanese
Monbukagakusho, and the Max Planck Society. The SDSS Web site is
\url{http://www.sdss.org/}.

The SDSS is managed by the Astrophysical Research Consortium (ARC) for
the Participating Institutions. The Participating Institutions are The
University of Chicago, Fermilab, the Institute for Advanced Study, the
Japan Participation Group, The Johns Hopkins University, Los Alamos
National Laboratory, the Max-Planck-Institute for Astronomy (MPIA), the
Max-Planck-Institute for Astrophysics (MPA), New Mexico State
University, University of Pittsburgh, Princeton University, the United
States Naval Observatory, and the University of Washington.

\bibliographystyle{apj}
\bibliography{MistyBentz}

\begin{thebibliography}{33}
\expandafter\ifx\csname natexlab\endcsname\relax\def\natexlab#1{#1}\fi

\bibitem[{{Abazajian} {et~al.}(2003)}]{dr1preprint}
{Abazajian}, K., {et~al.} 2003, \aj, in press (astro-ph/0305492)

\bibitem[{{Baldwin} {et~al.}(2003){Baldwin}, {Hamann}, {Korista}, {Ferland},
  {Dietrich}, \& {Warner}}]{2003ApJ...583..649B}
{Baldwin}, J.~A., {Hamann}, F., {Korista}, K.~T., {Ferland}, G.~J., {Dietrich},
  M., \& {Warner}, C. 2003, \apj, 583, 649

\bibitem[{{Barbaro} \& {Poggianti}(1997)}]{1997A&A...324..490B}
{Barbaro}, G., \& {Poggianti}, B.~M. 1997, \aap, 324, 490

\bibitem[{{Barkhouse} \& {Hall}(2001)}]{2001AJ....121.2843B}
{Barkhouse}, W.~A., \& {Hall}, P.~B. 2001, \aj, 121, 2843

\bibitem[{{Bolton} \& {Wall}(1970)}]{1970AuJPh..23..789B}
{Bolton}, J.~G., \& {Wall}, J.~V. 1970, Australian Journal of Physics, 23, 789

\bibitem[{{Boyle} {et~al.}(1990){Boyle}, {Fong}, {Shanks}, \&
  {Peterson}}]{1990MNRAS.243....1B}
{Boyle}, B.~J., {Fong}, R., {Shanks}, T., \& {Peterson}, B.~A. 1990, \mnras,
  243, 1

\bibitem[{{Chaffee} {et~al.}(1991){Chaffee}, {Foltz}, {Hewett}, {Francis},
  {Weymann}, {Morris}, {Anderson}, \& {MacAlpine}}]{1991AJ....102..461C}
{Chaffee}, F.~H., {Foltz}, C.~B., {Hewett}, P.~C., {Francis}, P.~A., {Weymann},
  R.~J., {Morris}, S.~L., {Anderson}, S.~F., \& {MacAlpine}, G.~M. 1991, \aj,
  102, 461

\bibitem[{{Chen} {et~al.}(2000){Chen}, {Han}, \& {He}}]{2000AcApS..20..366C}
{Chen}, Y., {Han}, J.-L., \& {He}, X.-T. 2000, Acta Astrophysica Sinica, 20,
  366

\bibitem[{{Hamann}(1999)}]{1999quco.conf..409H}
{Hamann}, F. 1999, in ASP Conf. Ser. 162: Quasars and Cosmology, 409

\bibitem[{{Hamann} \& {Ferland}(1999)}]{1999AR}
{Hamann}, F., \& {Ferland}, G. 1999, \araa, 37, 487

\bibitem[{{Hewett} {et~al.}(1991){Hewett}, {Foltz}, {Chaffee}, {Francis},
  {Weymann}, {Morris}, {Anderson}, \& {MacAlpine}}]{1991AJ....101.1121H}
{Hewett}, P.~C., {Foltz}, C.~B., {Chaffee}, F.~H., {Francis}, P.~J., {Weymann},
  R.~J., {Morris}, S.~L., {Anderson}, S.~F., \& {MacAlpine}, G.~M. 1991, \aj,
  101, 1121

\bibitem[{{Hewitt} \& {Burbidge}(1987)}]{1987ApJS...63....1H}
{Hewitt}, A., \& {Burbidge}, G. 1987, \apjs, 63, 1

\bibitem[{{Huang} \& {Usher}(1984)}]{1984ApJS...56..393H}
{Huang}, K.-L., \& {Usher}, P.~D. 1984, \apjs, 56, 393

\bibitem[{{Kennicutt}(1998)}]{Kennicutt1998}
{Kennicutt}, R.~C. 1998, \araa, 36, 189

\bibitem[{{La Franca} {et~al.}(1994){La Franca}, {Gregorini}, {Cristiani}, {de
  Ruiter}, \& {Owen}}]{1994AJ....108.1548L}
{La Franca}, F., {Gregorini}, L., {Cristiani}, S., {de Ruiter}, H., \& {Owen},
  F. 1994, \aj, 108, 1548

\bibitem[{{Leitherer} {et~al.}(1999){Leitherer}, {Schaerer}, {Goldader},
  {Delgado}, {Robert}, {Kune}, {de Mello}, {Devost}, \&
  {Heckman}}]{Leitherer1999}
{Leitherer}, C., {Schaerer}, D., {Goldader}, J.~D., {Delgado}, R.~M.~G.,
  {Robert}, C., {Kune}, D.~F., {de Mello}, D.~F., {Devost}, D., \& {Heckman},
  T.~M. 1999, \apjs, 123, 3

\bibitem[{{MacAlpine} {et~al.}(1977){MacAlpine}, {Lewis}, \&
  {Smith}}]{1977ApJS...35..203M}
{MacAlpine}, G.~M., {Lewis}, D.~W., \& {Smith}, S.~B. 1977, \apjs, 35, 203

\bibitem[{{Morton}(1991)}]{1991ApJS...77..119M}
{Morton}, D.~C. 1991, \apjs, 77, 119

\bibitem[{{Osmer}(1980)}]{1980ApJ...237..666O}
{Osmer}, P.~S. 1980, \apj, 237, 666

\bibitem[{{Osmer} {et~al.}(1994){Osmer}, {Porter}, \&
  {Green}}]{1994ApJ...436..678O}
{Osmer}, P.~S., {Porter}, A.~C., \& {Green}, R.~F. 1994, \apj, 436, 678

\bibitem[{{Osmer} \& {Smith}(1980)}]{1980ApJS...42..333O}
{Osmer}, P.~S., \& {Smith}, M.~G. 1980, \apjs, 42, 333

\bibitem[{{Perlman} {et~al.}(1998){Perlman}, {Padovani}, {Giommi}, {Sambruna},
  {Jones}, {Tzioumis}, \& {Reynolds}}]{1998AJ....115.1253P}
{Perlman}, E.~S., {Padovani}, P., {Giommi}, P., {Sambruna}, R., {Jones}, L.~R.,
  {Tzioumis}, A., \& {Reynolds}, J. 1998, \aj, 115, 1253

\bibitem[{{Richards} {et~al.}(2002)}]{2002AJ....123.2945R}
{Richards}, G.~T., {et~al.} 2002, \aj, 123, 2945

\bibitem[{{Schneider} {et~al.}(2002)}]{2002AJ....123..567S}
{Schneider}, D.~P., {et~al.} 2002, \aj, 123, 567

\bibitem[{{Shapley} {et~al.}(2003){Shapley}, {Steidel}, {Pettini}, \&
  {Adelberger}}]{2003ApJ...588...65S}
{Shapley}, A.~E., {Steidel}, C.~C., {Pettini}, M., \& {Adelberger}, K.~L. 2003,
  \apj, 588, 65

\bibitem[{{Simard-Normandin} {et~al.}(1981){Simard-Normandin}, {Kronberg}, \&
  {Button}}]{1981ApJS...46..239S}
{Simard-Normandin}, M., {Kronberg}, P.~P., \& {Button}, S. 1981, \apjs, 46, 239

\bibitem[{{Sirola} {et~al.}(1998){Sirola}, {Turnshek}, {Weymann}, {Monier},
  {Morris}, {Roth}, {Krzeminski}, {Kunkel}, {Duhalde}, \&
  {Sheaffer}}]{1998ApJ...495..659S}
{Sirola}, C.~J., {Turnshek}, D.~A., {Weymann}, R.~J., {Monier}, E.~M.,
  {Morris}, S.~L., {Roth}, M.~R., {Krzeminski}, W., {Kunkel}, W.~E., {Duhalde},
  O., \& {Sheaffer}, S. 1998, \apj, 495, 659

\bibitem[{{Steidel} {et~al.}(2003){Steidel}, {Adelberger}, {Shapley},
  {Pettini}, {Dickinson}, \& {Giavalisco}}]{2003ApJ...592..728S}
{Steidel}, C.~C., {Adelberger}, K.~L., {Shapley}, A.~E., {Pettini}, M.,
  {Dickinson}, M., \& {Giavalisco}, M. 2003, \apj, 592, 728

\bibitem[{{Stoughton} {et~al.}(2002)}]{2002AJ....123..485S}
{Stoughton}, C., {et~al.} 2002, \aj, 123, 485

\bibitem[{{V{\' e}ron-Cetty} \& {V{\' e}ron}(2001)}]{2001A&A...374...92V}
{V{\' e}ron-Cetty}, M.-P., \& {V{\' e}ron}, P. 2001, \aap, 374, 92

\bibitem[{{Vanden Berk} {et~al.}(2001)}]{2001AJ....122..549V}
{Vanden Berk}, D.~E., {et~al.} 2001, \aj, 122, 549

\bibitem[{{Visnovsky} {et~al.}(1992){Visnovsky}, {Impey}, {Foltz}, {Hewett},
  {Weymann}, \& {Morris}}]{1992ApJ...391..560V}
{Visnovsky}, K.~L., {Impey}, C.~D., {Foltz}, C.~B., {Hewett}, P.~C., {Weymann},
  R.~J., \& {Morris}, S.~L. 1992, \apj, 391, 560

\bibitem[{{York} {et~al.}(2000)}]{2000AJ....120.1579Y}
{York}, D.~G., {et~al.} 2000, \aj, 120, 1579

\end{thebibliography}

\clearpage

\begin{figure}
\figurenum{1}
\epsscale{0.5}
\plotone{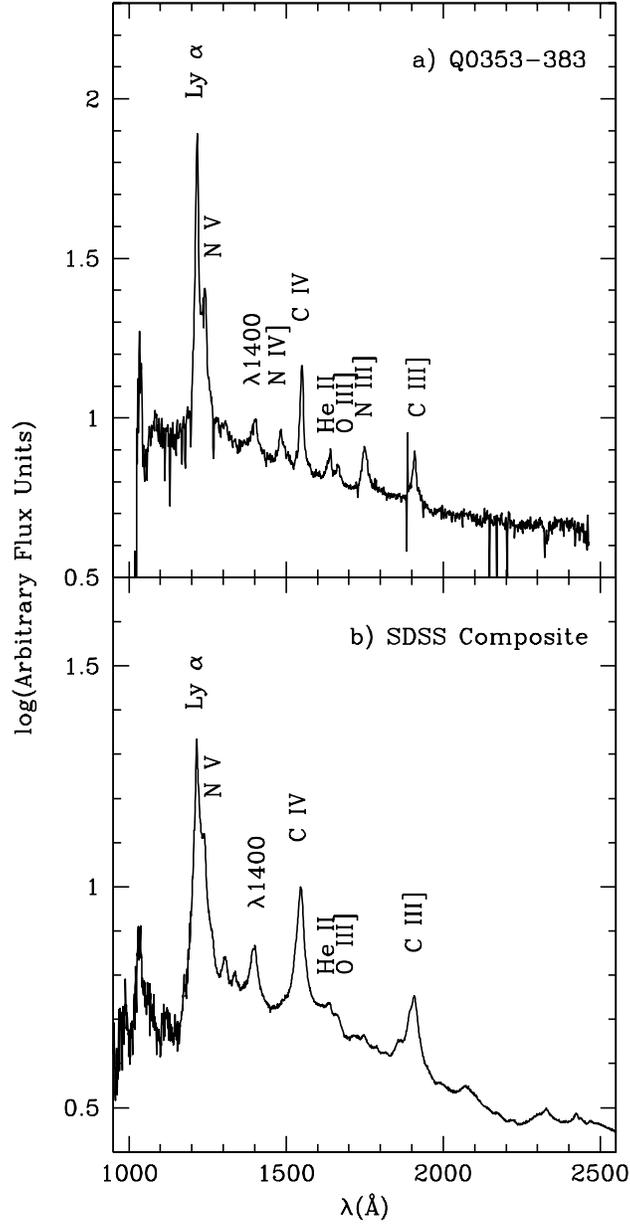}
\caption{Rest frame spectrum of Q0353-383 (Baldwin, private communication), 
        and the SDSS composite \citep[]{2001AJ....122..549V}, composed
        of 2204 QSO spectra from 66 spectroscopic plates.  Both spectra
        are plotted in semi-log format to enhance fine details.}
\end{figure}

\begin{figure}
\figurenum{2}
\epsscale{1}
\plotone{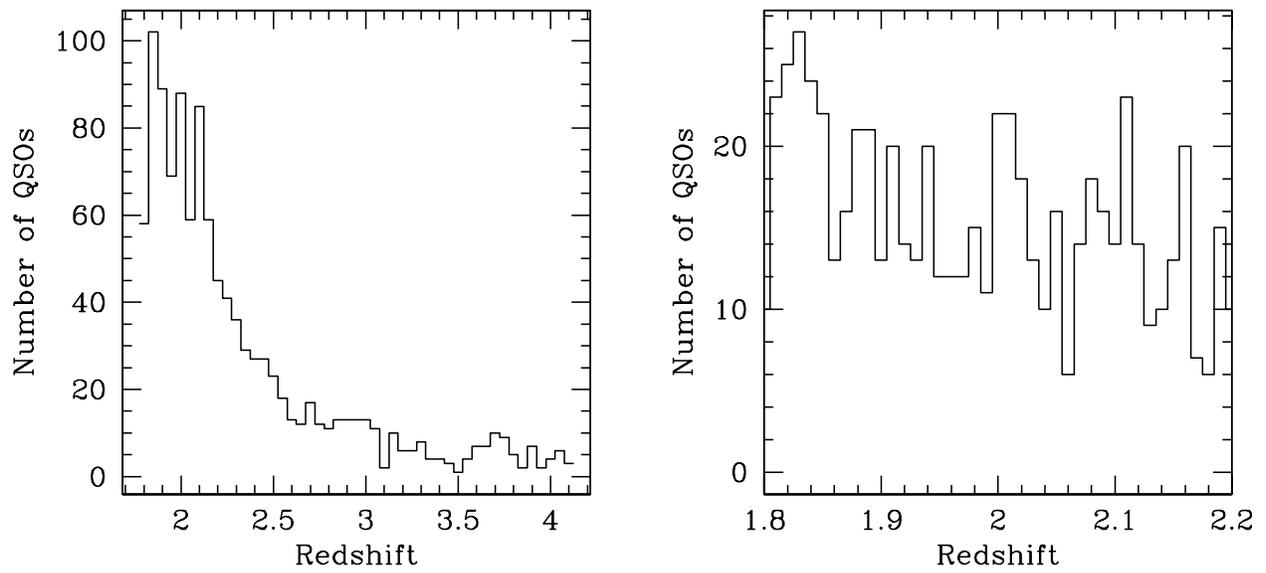}
\caption{Redshift distribution for the 1082 objects in the SDSS EDR with 
        $1.8 < z < 4.1$ and bin size $\Delta z = 0.05$, and distribution
        centered around $z = 2$ with bin size $\Delta z = 0.01$.}
\end{figure}

\begin{figure}
\figurenum{3}
\plotone{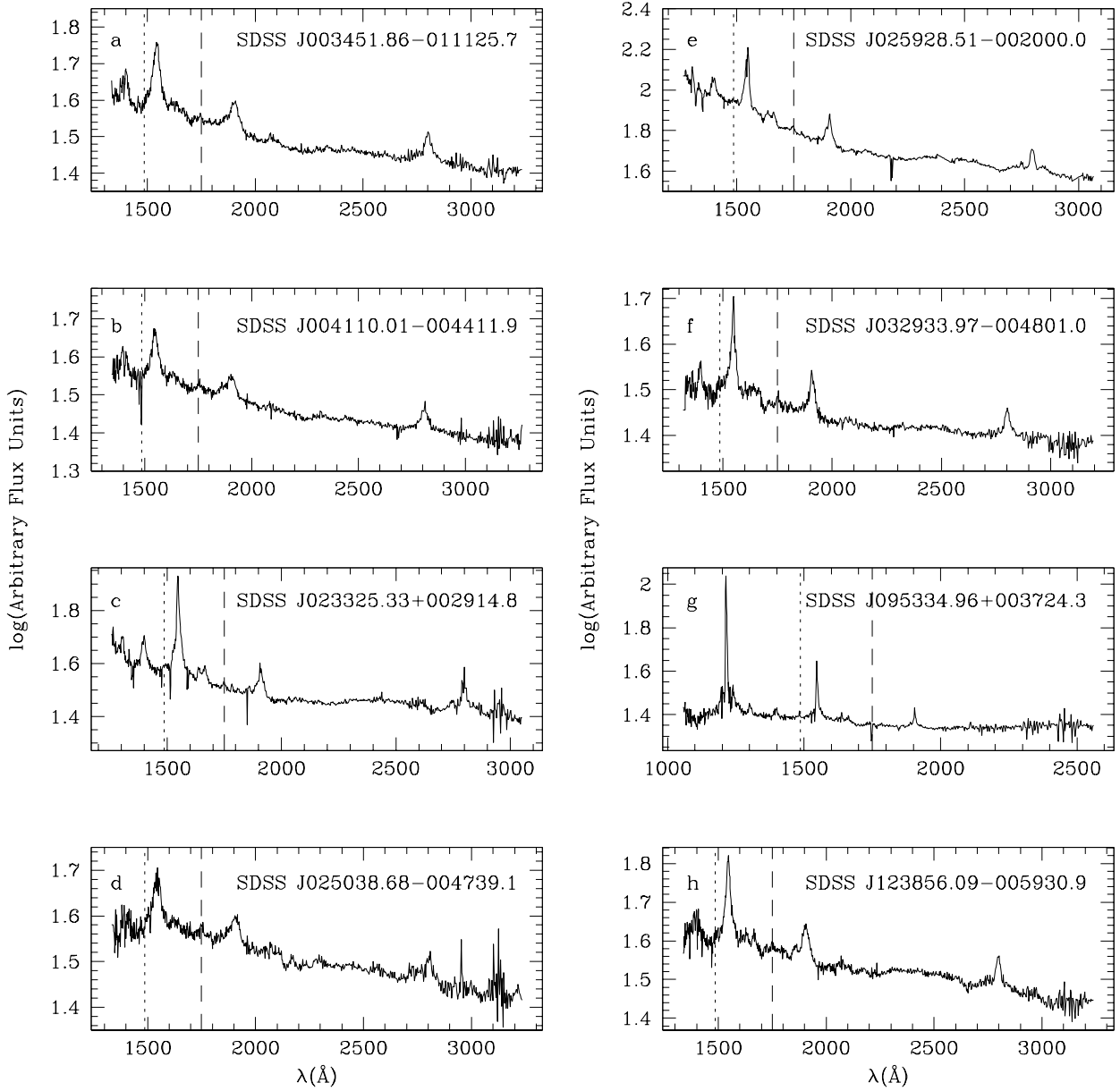}
\caption{Rest frame spectra of the 16 nitrogen enriched candidate QSOs 
         from the SDSS EDR database.  The spectra are smoothed with a
         bin of five pixels and plotted in semi-log format to enhance
         fine details.  The dotted line is at \ion{N}{4}] $\lambda$1486
         \AA\ and the dashed line is at \ion{N}{3}] $\lambda$1750 \AA.}
\end{figure}
\begin{figure}
\plotone{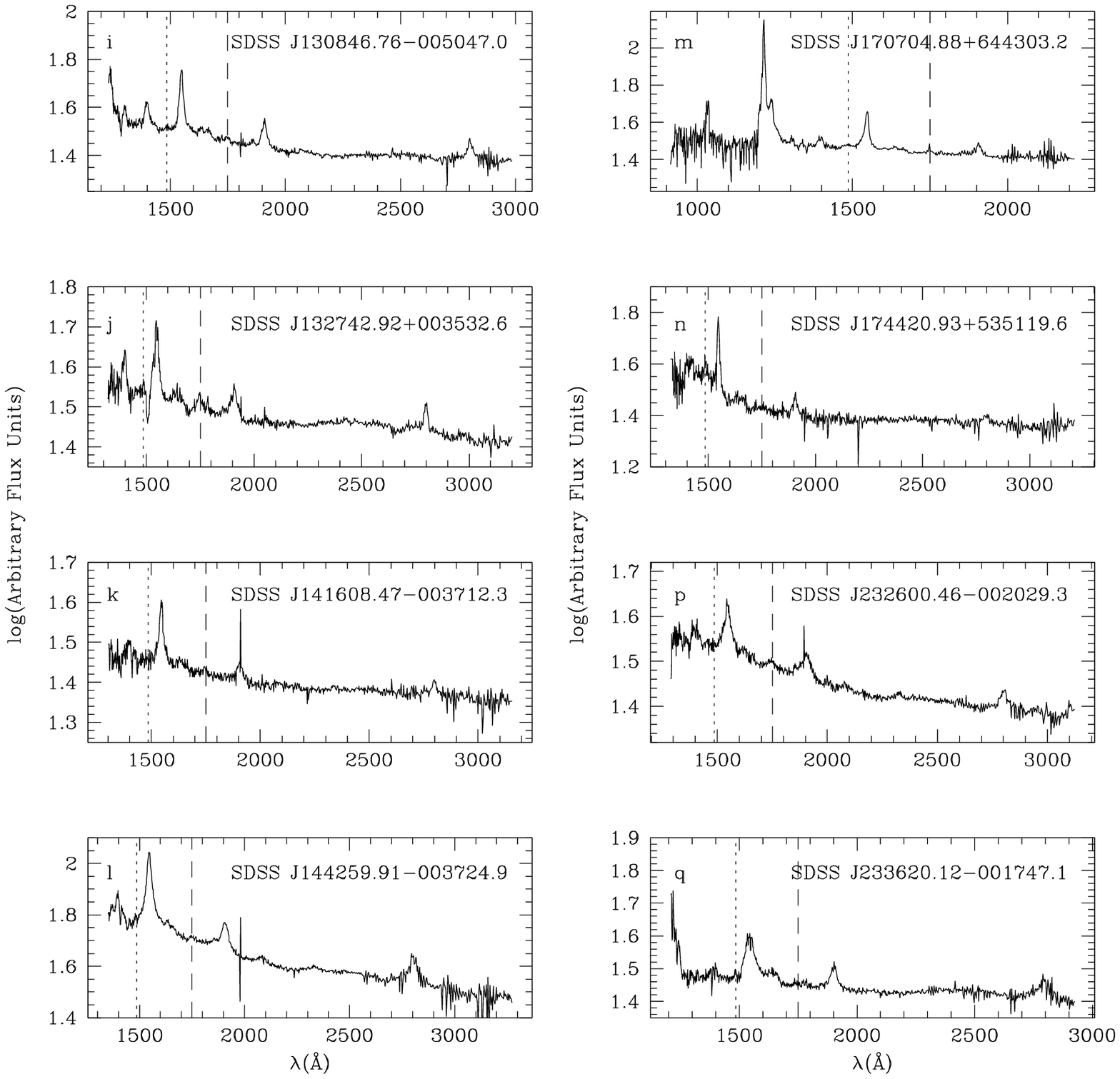}
\end{figure}

\begin{figure}
\figurenum{4}
\plotone{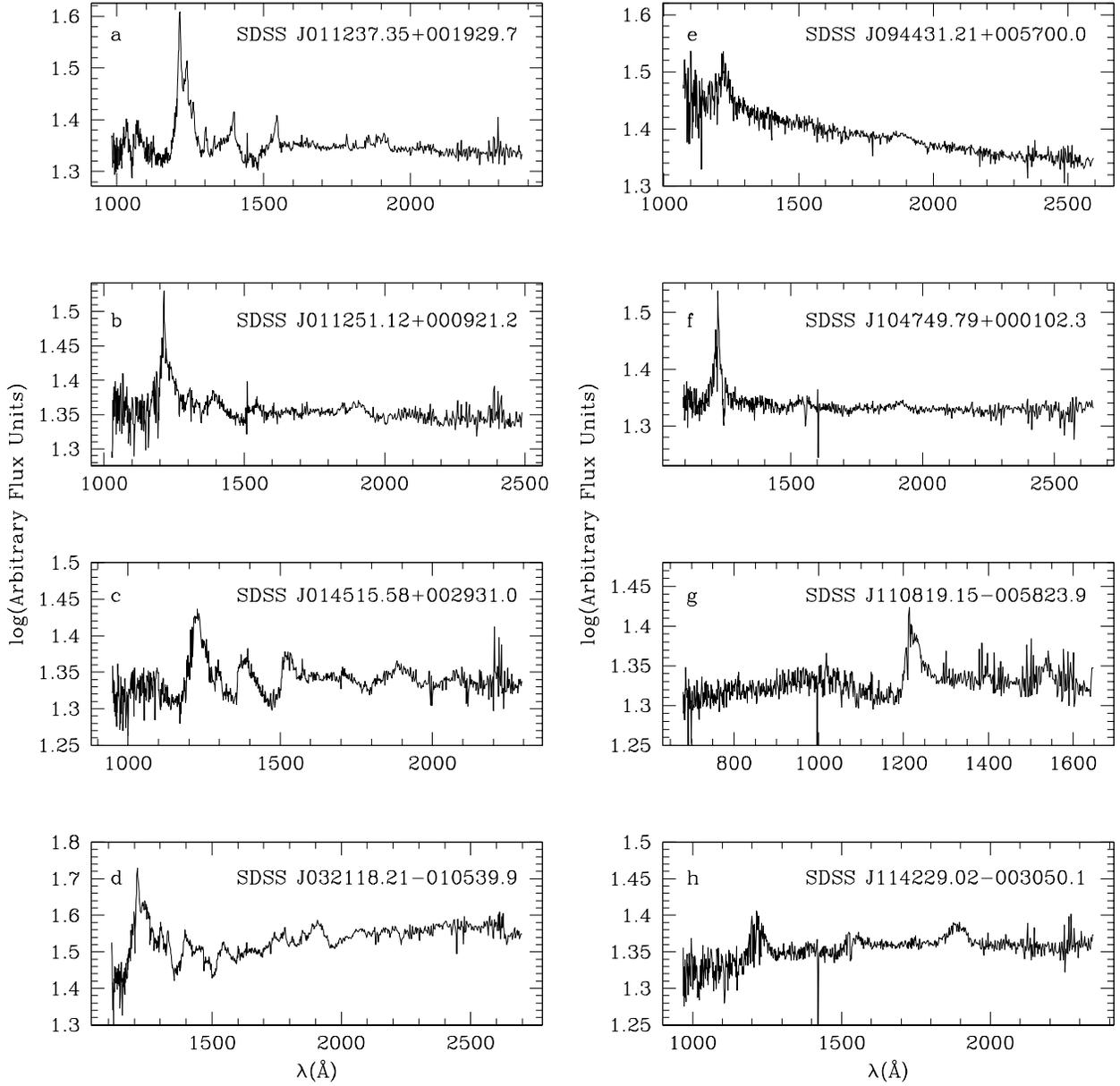}
\caption{Rest frame spectra of the 23 unusual quasars.  The spectra are 
         smoothed with a bin of five pixels and plotted in semi-log
         format to enhance fine details.}
\end{figure}
\begin{figure}
\plotone{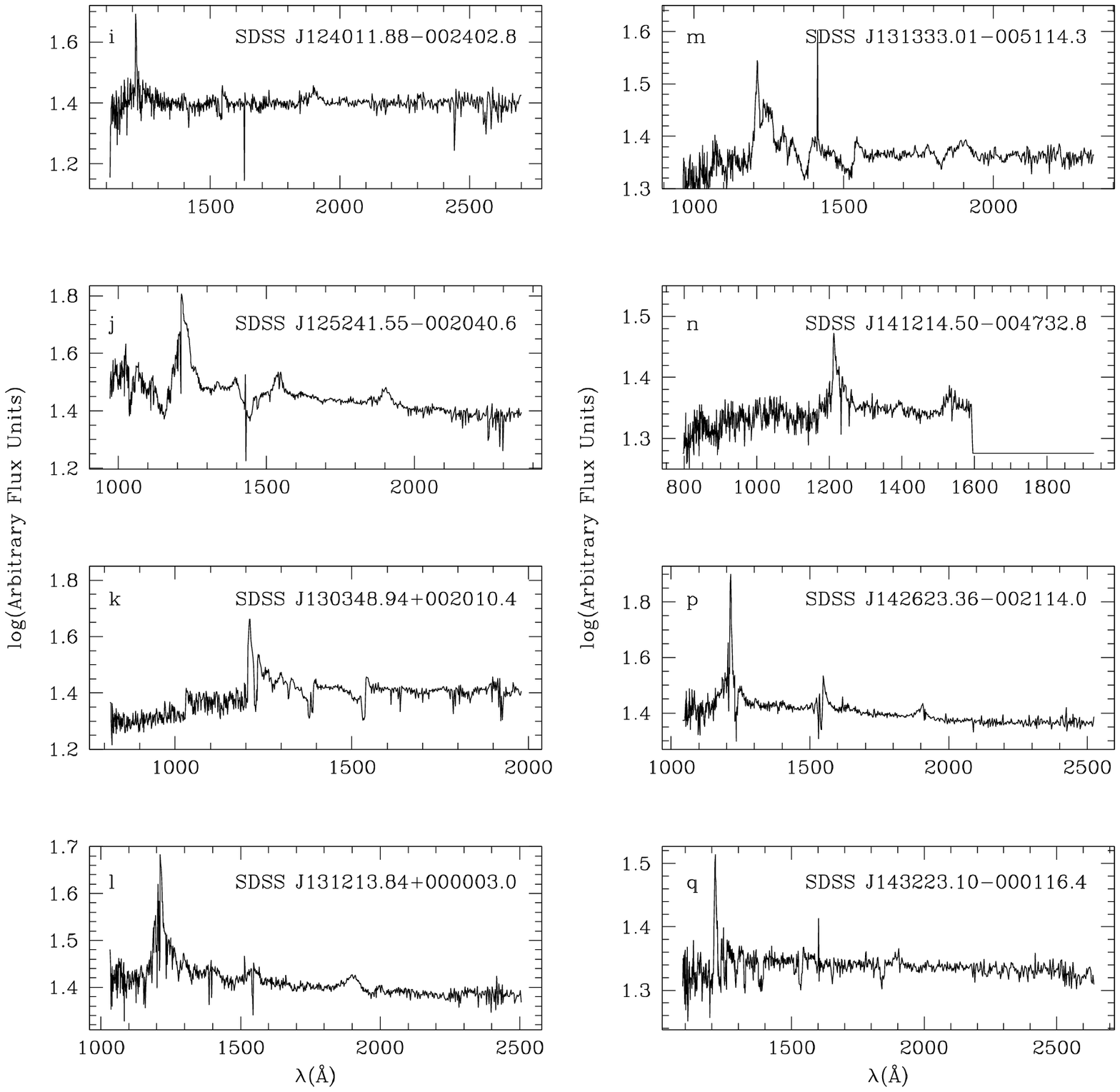}
\end{figure}
\begin{figure}
\plotone{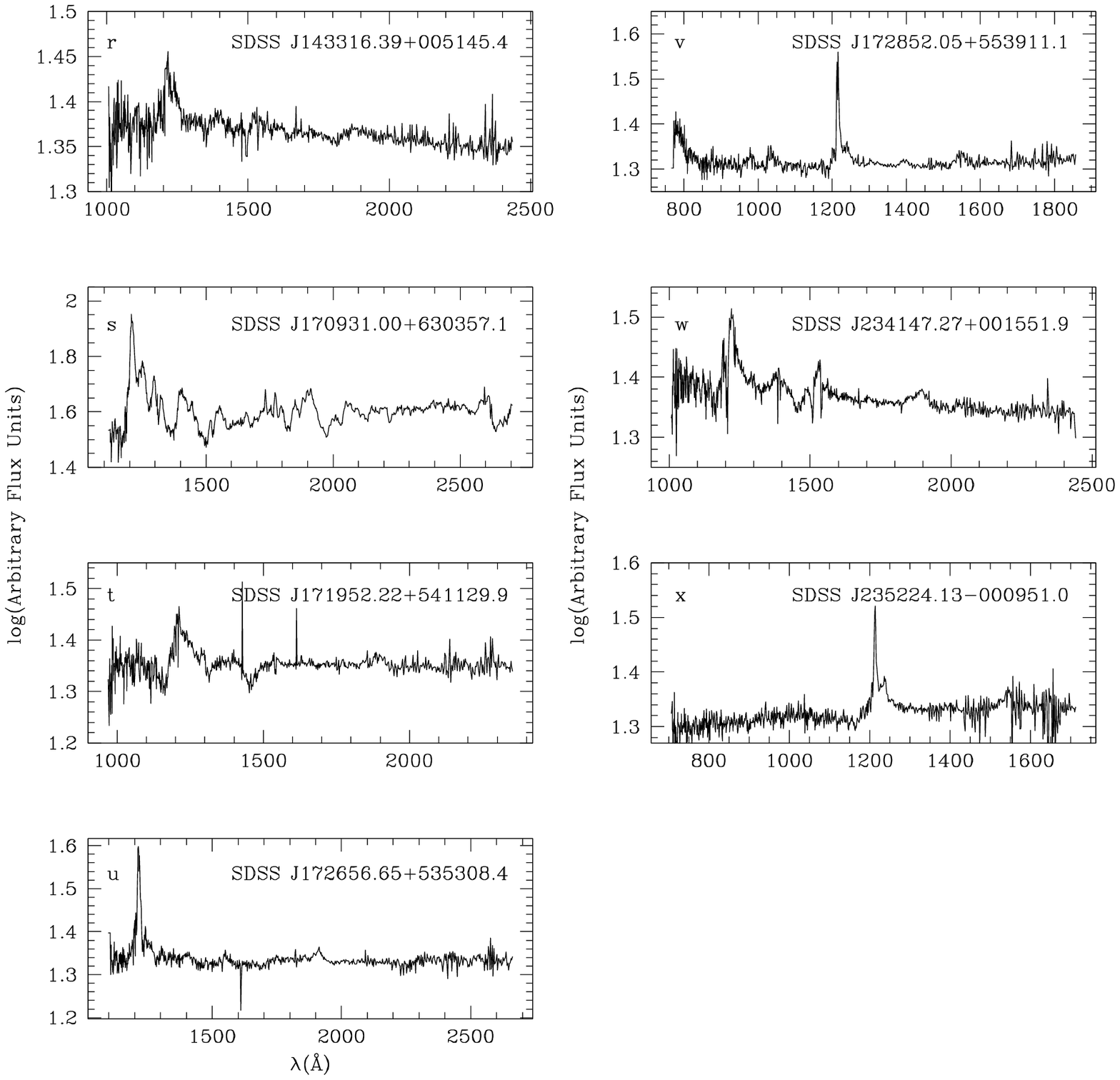}
\end{figure}

\begin{figure}
\figurenum{5}
\plotone{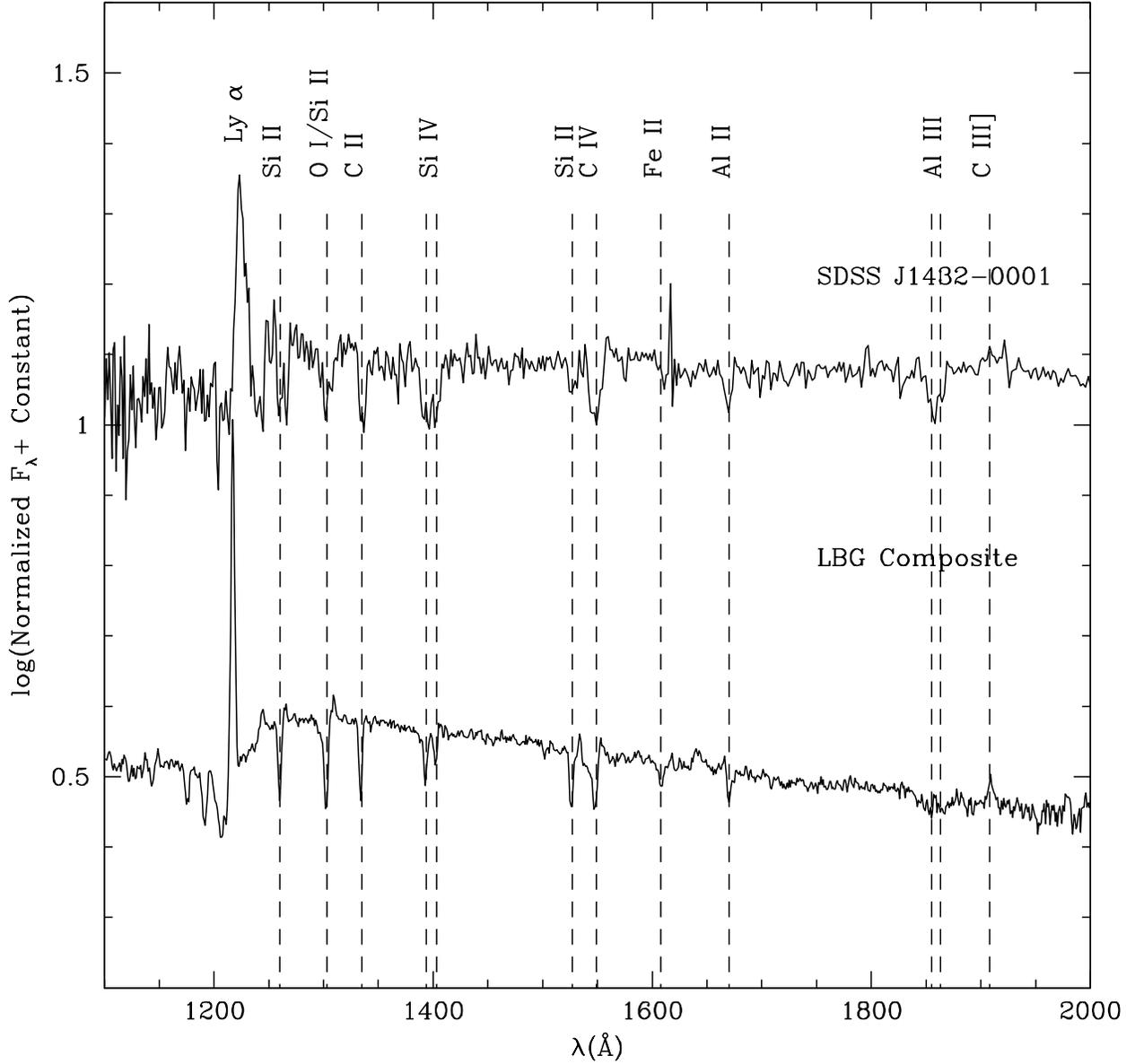}
\caption{Comparison of rest frame composite spectrum, made up of 811 
        individual Lyman break galaxies \citep{2003ApJ...588...65S}, and
        Sloan object SDSS J1432-0001.  The Sloan spectrum was smoothed
        with a bin of five pixels, and both spectra are plotted in
        semi-log space to enhance fine details.}
\end{figure}

\begin{deluxetable}{lccclll} 
\tabletypesize{\footnotesize}
\tablecolumns{7} 
\tablewidth{0pc}
\tablecaption{Properties of Nitrogen Enriched Candidate Quasars}
\tablehead{ 
\colhead{} & 
\colhead{} &
\multicolumn{2}{c}{i* Magnitude\tablenotemark{a}} & 
\colhead{} & 
\colhead{} &
\colhead{}\\ 
\cline{3-4} \\ 
\colhead{QSO} & 
\colhead{z\tablenotemark{a}} & 
\colhead{Obs.} & 
\colhead{Abs.} & 
\colhead{Add. Identifier} & 
\colhead{Notes\tablenotemark{b}} &
\colhead{References\tablenotemark{c}}} 
\startdata
SDSS J003451.86$-$011125.7	& 1.846	&18.13	& -27.07 & UM 259 &L3 &10, 11, 15\\
SDSS J004110.00$-$004411.8	& 1.819	&18.35	& -26.81 & & &\\
SDSS J023325.33+002914.9	& 2.014	&18.12	& -27.26 & & &\\
SDSS J025038.68$-$004739.1	& 1.845	&18.10	& -27.17 & US 3221 &RQ &8, 9, 15\\ 
SDSS J025928.52$-$002000.0	& 2.002	&16.98	& -28.54 & PKS 0256-005 &RL, FSRQ, L1 & 1, 2, 4, 7, 11, 12, 13, 15\\
SDSS J032933.97$-$004801.2	& 1.880	&18.55	& -26.85 & & \\
SDSS J095334.96+003724.5	& 2.606	&19.19	& -26.77 & & \\
SDSS J123856.10$-$005930.9	& 1.848	&17.95	& -27.25 & QNY4:41 &RD, L2 & 3, 5, 6, 15\\
SDSS J130846.76$-$005046.9	& 2.087	&18.85	& -26.60 & & \\
SDSS J132742.92+003532.6	& 1.876	&18.34	& -26.89 & & \\
SDSS J141608.47$-$003712.3	& 1.921	&18.98	& -26.34 & & \\
SDSS J144259.91$-$003724.9	& 1.818	&17.51	& -27.69 & LBQS 1440-0024 &RQ, OSV, L4 & 6, 14, 15, 16\\
SDSS J170704.88+644303.3	& 3.168	&18.42	& -27.90 & & \\
SDSS J174420.94+535119.5 	& 1.864	&19.63	& -25.62 & & \\
SDSS J232600.47$-$002029.3	& 1.946	&18.58	& -26.76 & & \\
SDSS J233620.13$-$001747.1	& 2.146	&18.35	& -27.19 & & \\
\enddata
\tablenotetext{a} {As determined by \citet[]{2002AJ....123..567S}, with 
                   $H_{0} = 50$, $\Omega_M = 1$, $\Omega_{\Lambda} = 0$,
                   and $\alpha_Q = 0.5$}
\tablenotetext{b} {\bf{Abbreviations}: \rm RL = Radio Loud; RD = Radio 
                   Detected; RQ = Radio Quiet; L1 = strong,
                   well-separated Ly $\alpha$ and \ion{N}{5}; L2 =
                   strong, separated Ly $\alpha$ and \ion{N}{5}; L3 =
                   broad Ly $\alpha$; L4= narrow Ly $\alpha$, \ion{N}{5}
                   indistinguishable; FSRQ = Flat Spectrum Radio Quasar;
                   OSV = Optically Strong Variable}
\tablenotetext{c} {This research has made use of the NASA/IPAC 
                  Extragalactic Database (NED) which is operated by the
                  Jet Propulsion Laboratory, California Institute of
                  Technology, under contract with the National
                  Aeronautics and Space Administration.}
\tablerefs{1. \citet{2001AJ....121.2843B}; 
           2. \citet{1970AuJPh..23..789B};
           3. \citet{1990MNRAS.243....1B};
           4. \citet{1991AJ....102..461C};
           5. \citet{2000AcApS..20..366C};
           6. \citet{1991AJ....101.1121H};
           7. \citet{1987ApJS...63....1H};
           8. \citet{1984ApJS...56..393H};
           9. \citet{1994AJ....108.1548L};
           10. \citet{1977ApJS...35..203M};
           11. \citet{1994ApJ...436..678O};
           12. \citet{1998AJ....115.1253P};
           13. \citet{1981ApJS...46..239S};
           14. \citet{1998ApJ...495..659S};
           15. \citet{2001A&A...374...92V};
           16. \citet{1992ApJ...391..560V}. }
\end{deluxetable}

\clearpage

\begin{deluxetable}{lccccccc}
\tabletypesize{\small}
\tablecolumns{8} 
\tablewidth{0pc}
\tablecaption{Measured Equivalent Widths \ion{N}{4}] $\lambda$1486 and \ion{N}{3}] $\lambda$1750 }
\tablehead{
\colhead{}  & 
\colhead{}  & 
\multicolumn{2}{c}{\ion{N}{4}]} & 
\multicolumn{2}{c}{\ion{N}{3}]} &
\multicolumn{2}{c}{CCC\tablenotemark{a}}\\
\cline{3-6}
\cline{7-8}\\
\colhead{QSO} & 
\colhead{SNR} & 
\colhead{$W_{obs}$ (\AA)\tablenotemark{b}} & 
\colhead{$\sigma$\tablenotemark{c}} & 
\colhead{$W_{obs}$ (\AA)\tablenotemark{b}} & 
\colhead{$\sigma$\tablenotemark{c}} &
\colhead{$\lambda 1486$ \AA} &
\colhead{$\lambda 1750$ \AA}} 
\startdata
Q0353$-$383\tablenotemark{d} & \nodata & 5.0 & \nodata & 9.0 & \nodata & 0.98 & 0.99 \\
SDSS J0034$-$0111 & 15 & \nodata & \nodata & 3.0     & 8.2   & 0.04 & 0.72\\
SDSS J0041$-$0044 & 14 & \nodata & \nodata & 5.0     & 12.8  & 0.01 & 0.75\\
SDSS J0233+0029	  & 15 & \nodata & \nodata & 3.0     & 8.2   & 0.34 & 0.85\\
SDSS J0250$-$0047 & 14 & \nodata & \nodata & 3.0     & 7.7   & 0.29 & 0.67\\
SDSS J0259$-$0020 & 39 & \nodata & \nodata & 1.0     & 7.1   & 0.80 & 0.91\\
SDSS J0329$-$0048 & 9  & \nodata & \nodata & 5.0     & 8.2   & 0.15 & 0.82\\
SDSS J0953+0037	  & 9  & 1.0     & 1.6     & \nodata \tablenotemark{e}  & \nodata & 0.58 & 0.02\\
SDSS J1238$-$0059 & 22 & \nodata & \nodata & 2.0     & 8.0   & 0.02 & 0.64\\
SDSS J1308$-$0050 & 7  & 1.0     & 1.3     & 3.5     & 4.5   & INDEF\tablenotemark{f} & 0.79\\
SDSS J1327+0035	  & 10 & 1.5     & 2.7     & 12.0    & 21.9  & 0.18 & 0.93\\
SDSS J1416$-$0037 & 7  & \nodata & \nodata & 4.0     & 5.1   & 0.03 & 0.65\\
SDSS J1442$-$0037 & 26 & \nodata & \nodata & 2.0     & 9.5   & 0.01 & 0.73\\
SDSS J1707+6443	  & 19 & \nodata \tablenotemark{g} & \nodata  & 3.0  & 10.4 & 0.61 & 0.57\\
SDSS J1744+5351   & 4  & 7.0     & 5.1     & 1.5     & 1.1 & 0.74 & 0.10\\
SDSS J2326$-$0020 & 11 & 1.0     & 2.0     & 4.0     & 8.0 & 0.00 & 0.68\\
SDSS J2336$-$0017 & 13 & 3.0     & 7.1     & 2.0     & 4.7 & 0.57 & 0.32\\
\enddata
\tablenotetext{a} {Cross-correlation coefficients for the 30 \AA\ 
                   windows as described in the text; auto-correlation
                   coefficients for Q0353-383}
\tablenotetext{b} {Guideline measurements, with errors of 0.5-1.0 \AA}
\tablenotetext{c} {Significance of detection, assuming a 30 \AA\ window}
\tablenotetext{d} {Values taken from \citet{2003ApJ...583..649B}}  
\tablenotetext{e} {Obstructed by [\ion{O}{1}] sky line}
\tablenotetext{f} {IRAF fit did not converge} 
\tablenotemark{g} {Obstructed by Fe emission}
\end{deluxetable}

\clearpage

\begin{deluxetable}{lcc}
\tablewidth{0pc}
\tablecolumns{3}
\tablecaption{Measured Equivalent Widths \ion{He}{2} $\lambda$1640 and \ion{O}{3}] $\lambda$1664}
\tablehead{
\colhead{} & 
\multicolumn{2}{c}{$W_{obs}$ (\AA)\tablenotemark{a}} \\
\cline{2-3} \\
\colhead{QSO} & 
\colhead{\ion{He}{2}} & 
\colhead{\ion{O}{3}]}}
\startdata
Q0353$-$383 & 4.0 & 2.0 \\ SDSS J0034$-$0111 & 3.6 & 3.0 \\ SDSS
J0233+0029 & 4.7 & 5.7 \\ SDSS J0259$-$0020 & 9.6 & 4.4 \\ SDSS
J0953$-$0037 & 1.6 & 1.0 \\ SDSS J1238$-$0059 & \nodata & 5.6 \\ SDSS
J1308$-$0050 & 2.5 & 2.5 \\ SDSS J1442$-$0037 & 2.0 & 0.6 \\
\enddata
\tablenotetext{a} {Guideline measurements, with errors of 0.5-1.0 \AA} 
\end{deluxetable}

\begin{deluxetable}{lccccl}
\tabletypesize{\footnotesize}
\tablewidth{0pc}
\tablecolumns{6}
\tablecaption{Properties of Quasars with Small 
              \ion{C}{4}/(Ly$\alpha$ + \ion{N}{5})}
\tablehead{
\colhead{} &
\colhead{} &
\colhead{} &
\multicolumn{2}{c}{i* Magnitude\tablenotemark{a}} &
\colhead{} \\ 
\cline{4-5} \\
\colhead{QSO} & 
\colhead{z\tablenotemark{a}} & 
\colhead{R\tablenotemark{b}} &
\colhead{Observed} & 
\colhead{Absolute} & 
\colhead{Characteristics}} 
\startdata
SDSS J011251.12+000921.2  	& 2.865	& 0.096 & 19.70	& -26.42 & strong, narrow Ly $\alpha$, well separated \ion{N}{5}\\
SDSS J011237.35+001929.7  	& 2.695	& 0.081 & 19.60	& -26.40 & broad and narrow Ly $\alpha$ components\\
SDSS J014515.58+002931.0	& 3.006	& 0.077 & 19.87	& -26.35 & BAL quasar\\
SDSS J032118.21$-$0010539.9	& 2.412	& 0.044 & 17.61	& -28.25 & BAL quasar\\
SDSS J094431.21+005700.0	& 2.550	& 0.081 & 19.30	& -26.75 & slope to blue and broad, weak emission lines\\
SDSS J104749.79+000102.3	& 2.479	& 0.066 & 19.99	& -25.85 & narrow, intrinsic \ion{C}{4} and \ion{N}{5} absorption\\
SDSS J110819.15$-$005823.9	& 4.560	& 0.047 & 19.89	& -27.19 & broad Ly $\alpha$\\
SDSS J114229.02$-$003050.0	& 2.927	& 0.007 & 19.50	& -26.64 & slope to red and broad, strong emission lines\\
SDSS J124011.88$-$002402.8	& 2.417	& 0.091 & 19.29	& -26.47 & narrow, intrinsic \ion{C}{4} and \ion{N}{5} absorption\\
SDSS J125241.55$-$002040.6	& 2.898	& 0.010 & 18.53	& -27.61 & BAL quasar\\
SDSS J130348.94+002010.4	& 3.655	& 0.057 & 18.72	& -27.87 & multiple P-Cygni lines (\ion{C}{4}, \ion{N}{5}, Si $\lambda$1400, etc.)\\
SDSS J131213.84+000003.0	& 2.681	& 0.043 & 18.89	& -27.10 & narrow, intrinsic \ion{C}{4} and \ion{N}{5} absorption\\
SDSS J131333.01$-$005114.3	& 2.949	& 0.077 & 19.16	& -27.00 & BAL quasar\\
SDSS J141214.50$-$004732.8	& 3.777	& 0.073 & 19.82 & -26.90 & narrow, intrinsic \ion{C}{4} and \ion{N}{5} absorption\\
SDSS J142623.36$-$002114.0	& 2.644	& 0.000 & 19.23	& -26.74 & narrow, intrinsic \ion{C}{4} and \ion{N}{5} absorption\\
SDSS J143223.10$-$000116.4	& 2.483	& 0.056 & 20.12	& -25.73 & high redshift star-forming galaxy\\
SDSS J143316.39+005145.4	& 2.774	& 0.096 & 19.61	& -26.45 & BALs and narrow, intrinsic \ion{C}{4} absorption\\
SDSS J170931.00+63.0357.1	& 2.402	& 0.034 & 17.31	& -28.44 & BAL quasar\\
SDSS J172656.65+535308.4	& 2.905	& 0.029 & 19.70	& -26.47 & BALs and narrow, intrinsic \ion{C}{4} absorption\\
SDSS J172852.05+553911.2	& 2.461	& 0.083 & 20.21	& -25.62 & broad and narrow Ly $\alpha$ components\\
SDSS J234147.27+001551.9	& 3.950	& 0.079 & 20.03	& -26.73 & strong, narrow Ly $\alpha$, well separated \ion{N}{5}\\
SDSS J235224.13$-$000951.0	& 2.769	& 0.042 & 19.24	& -26.81 & BALs and narrow, intrinsic \ion{C}{4} absorption\\
SDSS J235718.36+004350.3	& 4.340	& 0.073 & 19.87	& -27.09 & strong, narrow Ly $\alpha$, well separated \ion{N}{5}\\
\enddata
\tablenotetext{a} {As determined by \citet[]{2002AJ....123..567S}, 
                  with $H_{0} = 50$, $\Omega_M = 1$, $\Omega_{\Lambda} =
                  0$, and $\alpha_Q = 0.5$}
\tablenotetext{b} {Ratio of \ion{C}{4}/(Ly$\alpha$ + \ion{N}{5}), 
	           calculated using the equivalent widths of the
	           emission lines determined by the SDSS pipeline}
\end{deluxetable}

\end{document}